\documentclass[a4paper]{article}
\usepackage{amsmath,amssymb,graphicx}
\usepackage{subcaption}
\usepackage[fontsize=9.5pt]{scrextend}
\usepackage{multirow}
\usepackage{siunitx}
\usepackage{amsfonts}
\usepackage{bbm}
\usepackage{stmaryrd}
\usepackage{xcolor}
\usepackage{cite}
\usepackage{comment}
\usepackage[symbol]{footmisc}
\usepackage{INTERSPEECH2022}

\title{NASTAR: Noise Adaptive Speech Enhancement with Target-Conditional Resampling}
\name{\fontsize{11}{12}\selectfont Chi-Chang Lee$^{1,2*}$\footnote{equal contribution}, Cheng-Hung Hu$^{3*}$\footnotemark[1], Yu-Chen Lin$^{1,2}$, Chu-Song Chen$^{1,3}$, Hsin-Min Wang$^{3}$, Yu Tsao$^{2}$}

\address{
\fontsize{11}{12}\selectfont $^{1}$Department of Computer Science and Information Engineering, National Taiwan University, Taipei, Taiwan\\
\fontsize{11}{12}\selectfont $^{2}$Research Center for Information Technology Innovation, Academia Sinica, Taipei, Taiwan\\
\fontsize{11}{12}\selectfont $^{3}$Institute of Information Science, Academia Sinica, Taipei, Taiwan
}
\email{
r08922a27@csie.ntu.edu.tw,
n124345679976@citi.sinica.edu.tw,
f04922077@csie.ntu.edu.tw,
chusong@csie.ntu.edu.tw,
whm@iis.sinica.edu.tw,
yu.tsao@citi.sinica.edu.tw
}

\begin{document}

\maketitle
\begin{abstract}
For deep learning-based speech enhancement (SE) systems, the training-test acoustic mismatch can cause notable performance degradation. 
To address the mismatch issue, numerous noise adaptation strategies have been derived. 
In this paper, we propose a novel method, called noise adaptive speech enhancement with target-conditional resampling (NASTAR), which reduces mismatches with only one sample (one-shot) of noisy speech in the target environment.
NASTAR uses a feedback mechanism to simulate adaptive training data via a noise extractor and a retrieval model.
The noise extractor estimates the target noise from the noisy speech, called pseudo-noise.
The noise retrieval model retrieves relevant noise samples from a pool of noise signals according to the noisy speech, called relevant-cohort.
The pseudo-noise and the relevant-cohort set are jointly sampled and mixed with the source speech corpus to prepare simulated training data for noise adaptation.
Experimental results show that NASTAR can effectively use one noisy speech sample to adapt an SE model to a target condition.
Moreover, both the noise extractor and the noise retrieval model contribute to model adaptation.
To our best knowledge, NASTAR is the first work to perform one-shot noise adaptation through noise extraction and retrieval.
\end{abstract}

\begin{figure*}[h]
    \centering
    \begin{subfigure}[b]{0.39\textwidth}
        \centering
        \includegraphics[width=0.8\textwidth]{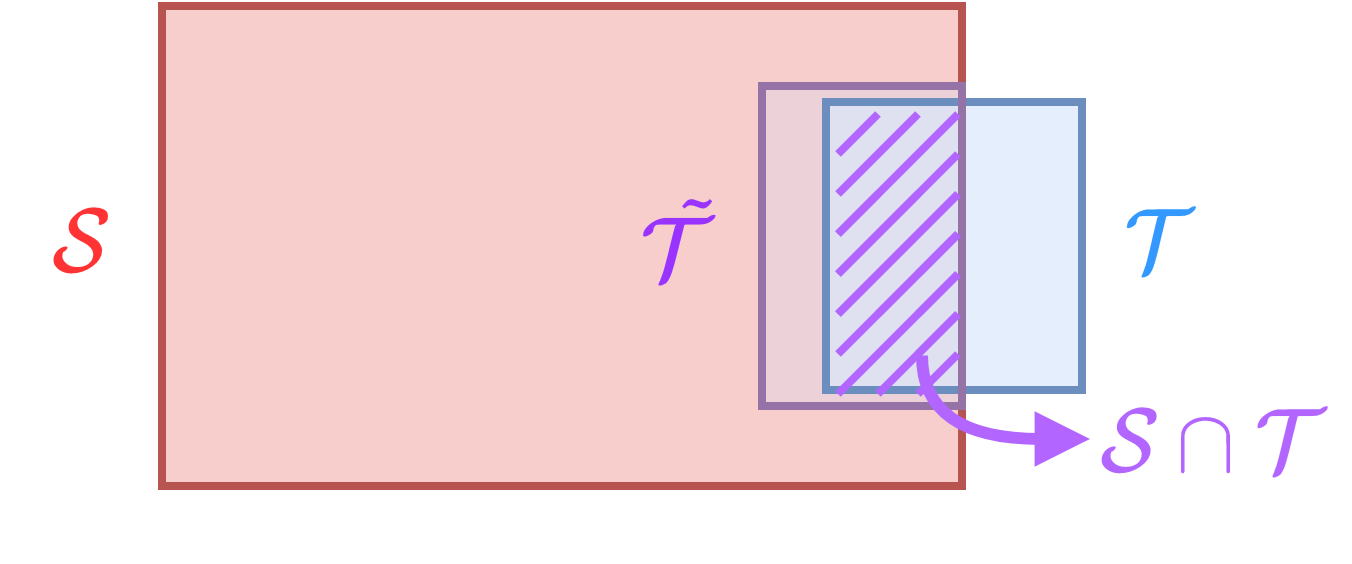}
        \subcaption{}
        \label{fig:intersection}
    \end{subfigure}
    \begin{subfigure}[b]{0.6\textwidth}
        \centering
        \includegraphics[width=1\textwidth]{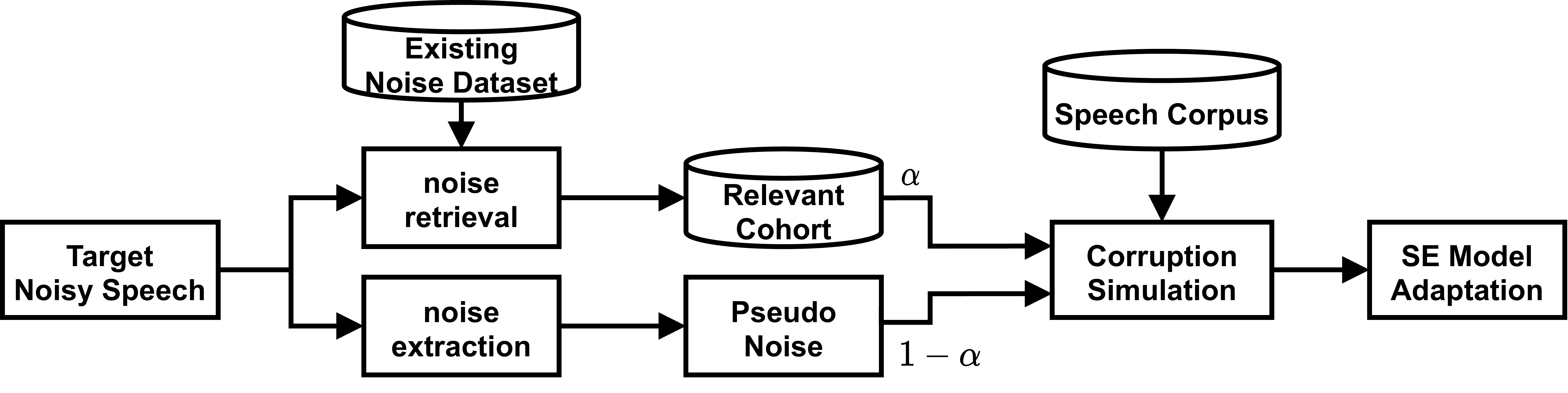}
        \subcaption{}
        \label{fig:flowchart}
    \end{subfigure}
    \setlength{\belowcaptionskip}{-10pt}
    \caption{(a) Illustration of the subset relation between the accessible source noise data set ($\mathcal{S}$), the target-similar noise set $\tilde{\mathcal{T}}$, and the true target noise set ($\mathcal{T}$); 
    (b) The flowchart of our NASTAR system. Given a query signal, the noise extractor directly estimates the pseudo-noise, and the retrieval model retrieves a relevant-cohort set. }
    \label{fig:overview}
    \vspace*{-0.1cm}
\end{figure*}

\section{Introduction}
\label{sec:intro}

Speech enhancement (SE) is commonly used as a pre-processor in speech-related applications, such as hearing aids~\cite{tan2019real, tinylstm}, automatic speech recognition~\cite{deepxi, 7952160, 10.1007/978-3-319-22482-4_11}, and speech emotion recognition~\cite{emotion2}.
Recently, various deep learning (DL) models have been used to formulate a regression function for SE~\cite{SE6, SE7, DAELu, CHLee, Liu2014ExperimentsOD, 7178061, SE_mul}.
Practically, we intend to collect a broad spectrum of noise types to train a unified DL-based SE model that performs well in diverse conditions. 
However, in real-world scenarios, the unified SE model inevitably encounters unseen noise types. 
If the training data do not entirely align with the testing distribution, a training-test mismatch occurs, limiting the achievable enhancement performance.

Thus far, numerous attempts have been made to reduce the training-test mismatch issue of SE.
To prevent mismatch under specific noise types, noise-aware strategies~\cite{CHEN20161, 8462126, noise_token, noise_aware3} were proposed to specify noise information into conditional embeddings to guide SE models to achieve better enhancement results. 
Meanwhile, ensemble learning strategies~\cite{7422753, 8170055, 6701888} were used in SE systems by preparing multiple component models, each of which is trained with a subset of the training data; during testing, the outputs of these component models are combined with a fusing/gating mechanism to dynamically handle noisy conditions.
Despite confirmed effectiveness in many tasks, noise-aware and ensemble learning SE systems may yield sub-optimal performance when the training data are not directly sampled for the target condition.
To handle the above issue, noise adaptation approaches have been proposed to handle mismatches by fine-tuning SE models to properly match target conditions.
For supervised adaptation methods, the pre-trained SE system is modified with the available noisy-clean paired signals~\cite{xu2014cross, wang2015transfer, 8369109, 7952122}.
For unsupervised adaptation methods, it is assumed that only noisy inputs are accessible. 
Several studies~\cite{liao2018noise, 8268914, lin2021unsupervised} applied domain adversarial training (DAT)~\cite{DAT} techniques to convert input speech signals to noise-invariant features for processing by the decoder of the SE model.
In addition, to avoid the catastrophic forgetting problem, Lee et al.~\cite{seril} proposed the SERIL system based on a regularization-based incremental learning strategy.
Meanwhile, multi-task learning approaches~\cite{SE6, bando2020adaptive} employ auxiliary tasks to provide positive effects for training adaptive SE models.
All of these existing strategies require batches of data for adaptation.
However, we usually only get a very small amount of data from the target environment.

\def\thefootnote{*}\footnotetext{These authors contributed equally to this work.}\def\thefootnote{\arabic{footnote}}
\def\thefootnote{$\dagger$}\footnotetext{The codes are availabel at https://github.com/ChangLee0903/NASTAR }

In this paper, we propose a novel method called noise adaptive speech enhancement with target-conditional resampling (NASTAR$^{\dagger}$) to effectively handle the mismatch problem with the least amount of target data. 
Instead of using batches of extra data, our work is the first to reduce the mismatch problem by filtering out out-of-target data.
Given one segment of noisy speech in the target environment, the NASTAR uses the sample to simulate the adaptation training data via a noise extractor and a retrieval model.
The noise extractor estimates the target noise from the noisy speech, termed pseudo-noise.
The noise retrieval model retrieves a relevant noise set, termed relevant-cohort, from an existing pool of noise signals based on the given noisy speech sample.
The pseudo-noise and the relevant-cohort set are used in a combined sampling scheme and then mixed with the source speech corpus to simulate training data for noise adaptation.
Compared to existing approaches, the main advantages of NASTAR include:

1) Data availability: NASTAR requires no additional target data beyond a given ``one-shot'' from the target environment. 
In noise adaptive SE research, this is the first work to re-utilize existing datasets instead of collecting additional data.

2) Promising performance: Experiment results confirm that NASTAR yields consistent and significant improvements (validated by the dependent t-test shown in Sec.~\ref{sec:results}) over a pre-trained model with few speech samples.

3) High training stability: NASTAR performs adaptation in a simple supervised manner, and the objective is committed to be aggressive towards handling the target noise type. We note that NASTAR has a more stable training process than the adaptive SE systems trained with multi-task objectives.

\section{Background and Motivation}
\label{sec:background}
Based on the nature of noise interference, given a clean utterance $s(t)$ and a noise signal $n(t)$, we can simulate the noisy signal by $s(t)+n(t)$ and thus prepare noisy-clean paired data of the target condition. 
In general, we assume that the recorded data from the target environment contain a mixture of speech and noise components. 
Thus, we can use a noise extractor to estimate the noise signal, termed pseudo-noise, from the noisy speech to simulate target noisy speech for model adaptation.
Nevertheless, the pseudo-noise may not be sufficient to cover the environment characteristics if the target noise is highly non-stationary, as there is only one segment.
Therefore, we need to leverage more target-similar noise signals for data simulation.
Although the prepared training data may cover a wide range of noise types, real-world SE systems may encounter few specific noise types not fully involved in all the training data, which inevitably leads to mismatches between training and test data, thereby limiting the SE performance.

To address the mismatch issue, this study investigates retrieving similar samples from the existing source noise dataset to enrich adaptation data.
Our main idea focuses on filtering out out-of-target data and finding ``what to train'' for the specific target environment to reduce the mismatch.
As shown in Fig.~\ref{fig:overview}(a), the proposed NASTAR system intends to collect the target-similar noise set $\tilde{\mathcal{T}}$ from the source noise dataset $\mathcal{S}$ and the pseudo-noise signal to sufficiently overlap with the true target noise set $\mathcal{T}$.
Since our resampling method selects noise signals from $\tilde{\mathcal{T}}$ instead of $\mathcal{S}$, this scheme is expected to reduce the mismatch between training and target noisy environments.
In this way, the combination of noise extractor and noise retrieval model can provide better noise adaptation results in a parallel training manner. 
Meanwhile, the NASTAR system only needs one target noisy speech.

In our experimental setup, we firstly constructed a custom noise set, which included $5$ different noise signals as different types.
Each type of the custom noise set corresponds to one target noise signal. 
We cut each target noise signal in half. 
The first half was used to contaminate a randomly selected test utterance as the accessible noisy speech (at SNR 0dB); 
the second half was used to contaminate other test utterances.

\section{The NASTAR System}
\label{sec:methodology}

\subsection{System overview}
\label{sec:sys}

Fig.~\ref{fig:flowchart} shows the overall flow of our proposed NASTAR system. 
First, we assume that a segment of target noisy speech (one-shot) is available and serves as the query signal. 
Next, the noise extractor estimates the noise from the query signal as pseudo-noise. 
Meanwhile, the noise retrieval model retrieves a relevant-cohort set from the source noise dataset based on the query signal. 
In this work, the relevant-cohort set consists of 250 noise signals closest to the query signal in terms of cosine similarity.
The collection of the pseudo-noise and the relevant-cohort set forms a noise pool that potentially covers the noise characteristics of the target condition.
By using a collaborative sampling mechanism, the pseudo-noise signal has the opportunity of ``1 - $\alpha$'', and each noise signal in the relevant-cohort set has the opportunity of ``$\alpha$/250'' to be selected to corrupt the clean speech. 
Finally, we adapt the pre-trained SE model to the target condition by using the sampled noise signals. $\alpha$ was set to 0.9.

\subsection{Noise extractor}
\label{sec:noise_extractor}

To construct a noise extractor, we used the DEMUCS model, which has been shown to yield state-of-the-art results on the music separation and SE tasks~\cite{demucs, demucs_stft}. 
Referred to \cite{demucs_stft}, we follow the same settings of the architecture and objective function to estimate the noise signal as pseudo-noise from the query signal.
The DEMUCS model consists of an encoder-decoder architecture with skip-connections and adopts a multi-resolution STFT objective function capturing the information of different time-frequency resolutions more effectively. 
The objective function is defined as:
\begin{equation}
\frac{1}{T}[||y-\hat{y}||_1+ \sum_{i=1}^{M} L_{stft}^{(i)}(y,\hat{y})],
\end{equation}
where $y$ and $\hat y$ are the target noise signal and the predicted noise signal, respectively; $M$ is the number of STFT losses; $L_{stft}$ is the addition of the spectral convergence loss $L_{sc}(y,\hat y)$ and the magnitude loss $L_{mag}(y,\hat y)$.
$L_{sc}(y,\hat y)$ and $L_{mag}(y,\hat y)$ are respectively calculated by Eq.~\ref{eq:lsc} and Eq.~\ref{eq:lmag}:
\begin{equation}
\label{eq:lsc}
L_{sc}(y,\hat y) = \frac{\||STFT(y)| - |STFT(\hat{y})|\|_F}{\||STFT(y)|\|_F},
\end{equation}
\begin{equation}
\label{eq:lmag}
L_{mag}(y,\hat y) =\frac{1}{T}\|\log|STFT(y)| - \log|STFT(\hat y)|\|_1. 
\end{equation}

\begin{figure}[h]
    \centering
    \begin{subfigure}[b]{0.135\textwidth}
        \includegraphics[width=\textwidth]{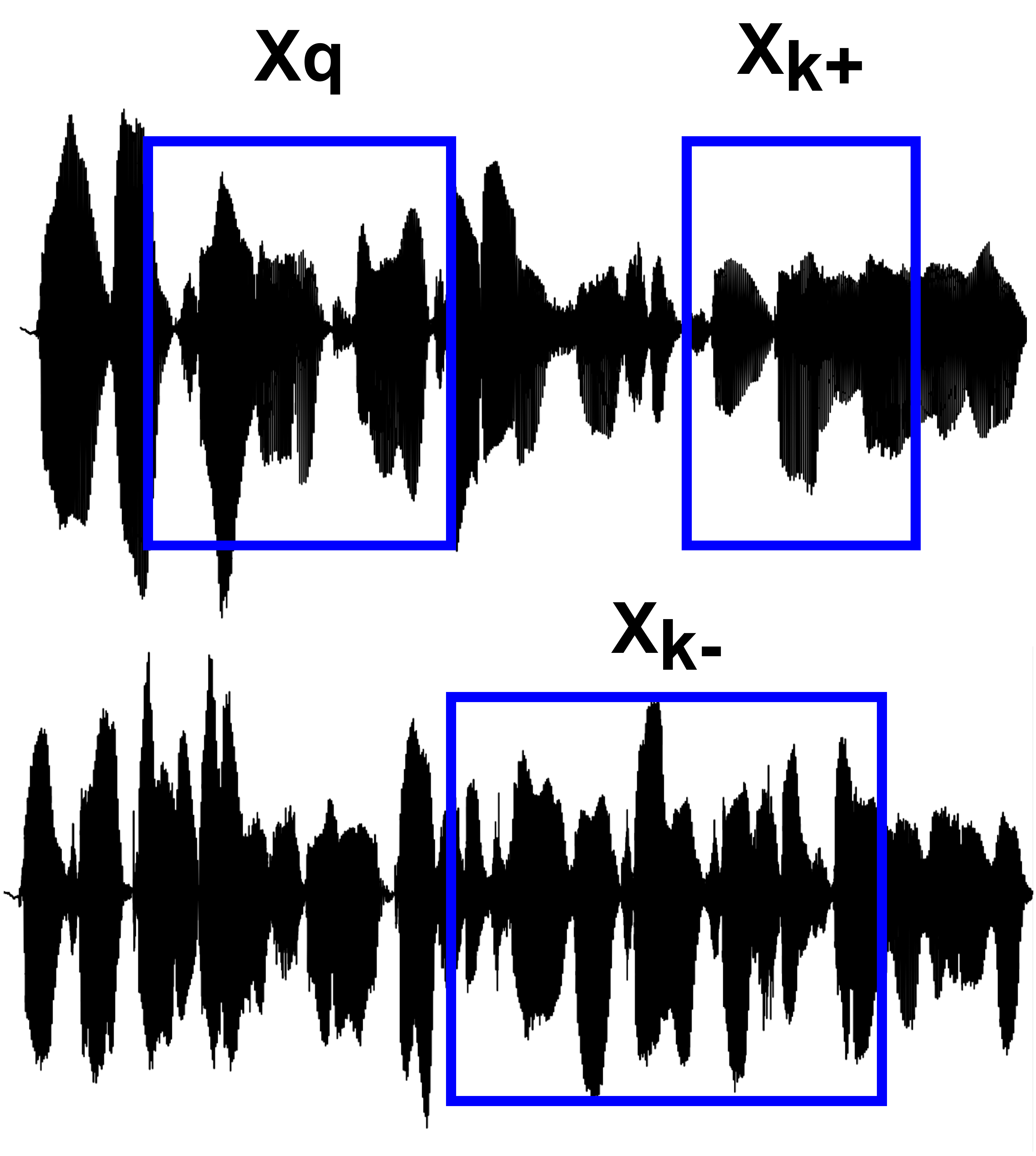}
        \label{fig:contrastive_learning a}
    \end{subfigure}
    \begin{subfigure}[b]{0.3\textwidth}
        \includegraphics[width=\textwidth]{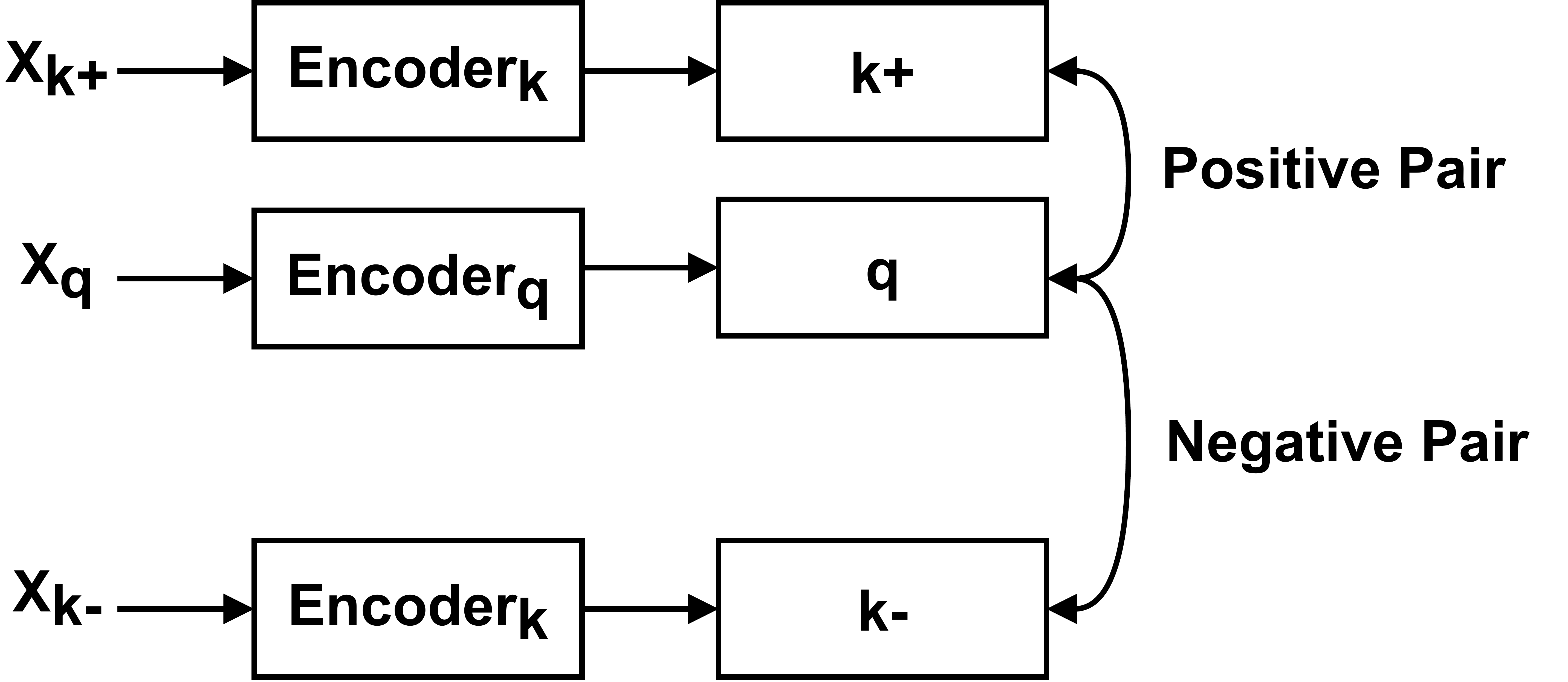}
        \label{fig:contrastive_learning b}
    \end{subfigure}
    \vspace*{-0.1cm}
    \caption{Formation of positive pairs and negative pairs for training the noise retrieval model.
    }
    \label{fig:contrastive_learning}
    \vspace*{-0.2cm}
\end{figure}

\subsection{Noise retrieval model}
\label{sec:contrastive}

Given a target noisy speech as the query signal, a noise retrieval model aims to determine whether a compared noise signal is relevant to the query signal. 
The key lies in a good representation of the query and compared signals. In the representation space, the relevant signals are closer to the query than the irrelevant signals. 
In addition, noise signals are diverse, complex, and thus it is challenging to define a noise type as a specific class. 

Some studies~\cite{sv_mi, sv_pcl} have explored self-supervised contrastive learning for speaker verification as an unsupervised learning framework. 
In~\cite{sv_pcl}, the commonly used contrastive learning frameworks SimCLR~\cite{simCLR}, MoCo~\cite{MoCo}, and ProtoNCE~\cite{pcl} are compared for extracting speaker representations. 
To learn discriminative noise representations, we refer to these frameworks and design a pretext task to make paired data to train our noise retrieval model, as shown in Fig.~\ref{fig:contrastive_learning}. 
Our pretext task used to produce paired segments from noise signals include truncation, duplication, and shifting operations. 
Each sampled noise signal is randomly processed into segments, and each segment ranges from 24,000 to 80,000 sample points.
Since different segments in the same noise signal have similar characteristics, these segments can be considered to be relevant to each other. 
As shown in Fig.~\ref{fig:contrastive_learning}, given a segment (the query segment $x_q$) excerpted from a noise signal, we can take another segment (the positive segment $x_{k^+}$ to be compared) from the same noise signal to form a positive pair.
Meanwhile, we assume that any two noise signals come from different noise conditions. Therefore, given a segment (the query segment $x_q$), we can randomly select a segment from another noise signal (the negative segment $x_{k^-}$ to be compared) to form a negative pair. 
In addition, each segment being compared has a 50\% chance of being mixed with a speech signal at a random SNR level (from -8dB to 8dB in 2dB steps). 
With these noisy speech segments involving both noise and speech components, our noise retrieval model can learn to filter out non-noise (speech) components. 
Therefore, at inference time, we can directly use ``noisy speech'' as the query signal to find those acoustically similar noise samples to construct the relevant-cohort set.

In training, we combine the negative pair usage mechanisms of the SimCLR~\cite{simCLR} and MoCo~\cite{MoCo} frameworks. 
In the early stage, in order to stabilize the training, we adopt the SimCLR method to generate negative pairs from the segments in a batch. 
After 5,000 steps, we follow MoCo to use the queue to access the past embeddings, and generate negative pairs from the past embeddings and the current embeddings. 
For a mini-batch, there are two embedding sets $\mathcal{B}_q$ and $\mathcal{B}_k$ generated by the query encoder network $f_q$ and the key encoder network $f_k$, respectively, and a queue $\mathcal{N}$ that stores embeddings in previous steps. 
The loss function is defined as:
\begin{equation}{
    l_{q, k^+} = -\log{\frac{\exp(\phi(q, k^+)/\tau)}{\sum\limits_{k^- \in \mathcal{B}_q \cup \mathcal{B}_k\cup\mathcal{N}}\mathbbm{1}_{[k^-\neq k^+]}\exp(\phi(q, k^-)/\tau)}},
    \label{eq:contrastive_learning}
}\end{equation}
where $\phi$ is the similarity function, and $\tau$ denotes the temperature parameter. 
The query embedding $q$ is obtained by $q = f_q(x_q)$, and $x_q$ is the query segment. 
The positive key embedding $k^+$ is obtained by $k^+ = f_k(x_k^+)$, and $x_k^+$ is a positive key segment. 
For the negative key embedding $k^-$ coming from $\mathcal{B}_q$ or $\mathcal{B}_k$, $k^-$ is obtained by $k^- = f_q(x_k^-)$ or $k^- = f_k(x_k^-)$, and $x_k^-$ is a negative key segment. 
The negative key embedding $k^-$ can also be sampled from the previous embeddings $\mathcal{N}$.
Given an embedding pair $(q,k) \in \mathbb{R}^d$, we use cosine similarity in the similarity function $\phi$.

Our noise retrieval model consists of three bidirectional LSTM layers and one cascaded MLP projection layer.
Denoting the parameters of $f_k$ as $\theta_k$ and those of $f_q$ as $\theta_q$, we update $\theta_k$ by $\theta_k =  \mu\theta_k + (1 - \mu)\theta_q$, where $\mu$ is the momentum coefficient.
To stabilize training in the early stage, those negative keys stored in $\mathcal{N}$ are not sampled until 5,000 steps. 

\begin{table*}[ht]
    \vspace*{-0.1cm}
    \caption{Average $\text{PESQ}_\text{nb}$, STOI, and SI-SDR scores of different models. The red and orange numbers denote the first and second place results for each condition, respectively.}
    \vspace*{-0.5cm}
    \label{tab:exp_SE}
    \setlength{\tabcolsep}{3pt}
    \fontsize{7.2}{9}\selectfont
    \center
    \begin{tabular}{|l|ccc|ccc|ccc|ccc|ccc|} 
\hline
 & \multicolumn{3}{c|}{ACVacuum} & \multicolumn{3}{c|}{Babble} & \multicolumn{3}{c|}{CafeRestaurant} & \multicolumn{3}{c|}{Car} & \multicolumn{3}{c|}{MetroSubway} \\ 
\hline
method & STOI & PESQ$_{\text{nb}}$ & SI-SDR & STOI & PESQ$_{\text{nb}}$ & SI-SDR & STOI & PESQ$_{\text{nb}}$ & SI-SDR & STOI & PESQ$_{\text{nb}}$ & SI-SDR & STOI & PESQ$_{\text{nb}}$ & SI-SDR \\ 
\hline
NOISY & 0.8407 & 1.9671 & 10.8404 & 0.8469 & 2.2791 & 12.3311 & 0.8955 & 2.7286 & 10.8916 & 0.8706 & 2.4135 & 11.9138 & 0.9359 & 3.3332 & 21.2817 \\
\hline
PTN & 0.8815 & 2.8496 & 17.6623 & 0.8862 & 2.8116 & 16.4004 & 0.9199 & 3.3711 & 17.5752 & 0.925 & 3.4061 & 16.8882 & 0.9488 & 3.8564 & 21.2581 \\
\hline
DAT$_{\text{one}}$ & 0.8823 & 2.8694 & 17.8753 & 0.8897 & 2.8017 & 16.9236 & 0.9215 & 3.3794 & 17.4182 & 0.9276 & 3.4170 & 17.2144 & 0.9504 & 3.8169 & 22.3700 \\
DAT$_{\text{full}}$ & 0.8712 & 2.5442 & 15.5812 & 0.884 & 2.7615 & 16.1574 & 0.9094 & 3.2511 & 16.7324 & 0.9089 & 3.0483 & 15.9469 & 0.9359 & 3.7047 & 20.5600 \\ 
NASTAR & \textcolor{red}{{0.8929}} & \textcolor{red}{{2.9482}} & 18.2684 & \textcolor{red}{{0.8916}} & \textcolor[rgb]{1,0.647,0}{{2.8655}} & \textcolor{red}{{17.2072}} & \textcolor{red}{{0.9244}} & \textcolor[rgb]{1,0.647,0}{{3.428}} & \textcolor{red}{{18.6476}} & \textcolor{red}{{0.930}} & \textcolor[rgb]{1,0.647,0}{{3.4602}} & \textcolor[rgb]{1,0.647,0}{{18.0286}} & \textcolor{red}{{0.9524}} & \textcolor{red}{{3.9121}} & 22.3703 \\ 
EXTR & 0.8914 & 2.9047 & \textcolor[rgb]{1,0.647,0}{{18.6487}} & 0.8771 & 2.6580 & 15.9827 & 0.9144 & 3.2114 & 14.9499 & 0.9113 & 3.1036 & 17.3962 & 0.9439 & 3.7233 & 21.5684 \\
GT & \textcolor[rgb]{1,0.647,0}{{0.8929}} & 2.9062 & \textcolor{red}{{18.7776}} & 0.8779 & 2.6576 & 16.1166 & 0.9167 & 3.2485 & 16.3388 & 0.9106 & 3.0888 & 17.2334 & 0.9457 & 3.7811 & \textcolor{red}{{23.3171}} \\
ALL & 0.8905 & 2.9330 & 18.2512 & 0.8903 & 2.8260 & 16.8602 & 0.9225 & 3.4186 & \textcolor[rgb]{1,0.647,0}{{18.4044}} & \textcolor[rgb]{1,0.647,0}{{0.9287}} & 3.4225 & 17.4977 & 0.9504 & 3.8201 & 22.0051 \\
RETV & 0.8883 & \textcolor[rgb]{1,0.647,0}{{2.9469}} & 18.1033 & \textcolor[rgb]{1,0.647,0}{{0.8913}} & \textcolor{red}{{2.8672}} & \textcolor[rgb]{1,0.647,0}{{17.1955}} & \textcolor[rgb]{1,0.647,0}{{0.9244}} & \textcolor{red}{{3.438}} & 18.3673 & 0.9282 & \textcolor{red}{{3.4949}} & \textcolor{red}{{18.2736}} & \textcolor[rgb]{1,0.647,0}{{0.9524}} & \textcolor[rgb]{1,0.647,0}{{3.8879}} & {\textcolor[rgb]{1,0.647,0}{22.7825}} \\
\hline
OPT & 0.9017 & 3.0561 & 18.9691 & 0.9111 & 3.0662 & 18.5652 & 0.9348 & 3.5684 & 19.4992 & 0.9364 & 3.5641 & 19.4694 & 0.9568 & 3.9346 & 26.5219 \\
\hline
    \end{tabular}
    \vspace*{-0.2cm}
\end{table*}

\section{Experiments}
\label{sec:exp}

\subsection{Experimental setup and implementation details}
\label{exp:implementation}

The datasets used in the experiments include: (1) the DNS-Challenge~\cite{DNS} noise dataset consisting of 65,303 background and foreground noise samples; (2) the Librispeech-360~\cite{panayotov2015librispeech} corpus consisting of 104,014 utterances; (3) the Voice Bank Corpus (VBC)~\cite{panayotov2015librispeech} consisting of 11,572 utterances of 28 speakers; and (4) the custom noise set consisting of five noise signals of five common noise types, namely \textit{AC/Vacuum}, \textit{Babble}, \textit{CafeRestaurant}, \textit{Car}, and \textit{MetroSubway}, selected from FreeSound~\cite{freesound}.

The NASTAR system consists of four modules: the noise extractor, the noise retrieval model, the pre-trained SE model, and the adapted SE model.
To train the noise extractor and the pre-trained SE model, the clean speech utterances from Librispeech-360 were contaminated by randomly sampled noise signals from DNS-Challenge at five SNR levels (from 0dB to 12dB in 3dB steps). 
We prepared noisy-noise and noisy-speech pairs to train the noise extractor and the pre-trained SE model, respectively. 
The models were trained by the Adam optimizer with $\beta_1$ = 0.9 and $\beta_2$ = 0.999, a learning rate of 0.0002, and a batch size of 8 with 500,000 steps. 
The noise extractor and the pre-trained SE model shared the same architecture and hyper-parameter settings.
To train the noise retrieval model, the training data were prepared from DNS-Challenge and Librispeech-360.
The process of preparing positive and negative pairs is presented in Sec.~\ref{sec:contrastive} and Fig.~\ref{fig:contrastive_learning}. 
The noise retrieval model was trained by contrasting positive and negative pairs. 
We used the Adam optimizer with $\beta_1$ = 0.9 and $\beta_2$ = 0.999, a learning rate of 0.00025, and a batch size of 256 to train the model for 100,000 steps.
The hyper-parameters were set as follows, the temperature coefficient $\tau$ = 0.1, the momentum coefficient $\mu$ = 0.9, and the queue size $|\mathcal{N}|$ = 32,768.
To train the adapted SE model, the pseudo-noise and the relevant-cohort set were derived based on the given query signal.
The pre-trained SE model served as the initial checkpoint. 
Each clean utterance from Librispeech-360 was mixed with a target-similar noise signal sampled from the pseudo-noise and relevant-cohort at 7 SNR levels (from -4dB to 8dB in 2dB steps). 
The SE model was updated by the Adam optimizer with $\beta_1$ = 0.9 and $\beta_2$ = 0.999, a learning rate of 0.0001, and a batch size of 8 with 20,000 steps.

For each noise signal in the custom noise set, we used its first half and a test speech from the test-clean set of Librispeech to form the 0dB query noisy speech and accordingly trained the adaptive SE model. 
That is, there are five adaptation models, each for one noise type. 
For evaluation, we used the clean speech of the VBC corpus as our test set.
For each noise condition, each clean utterance was separately mixed with the second half of the corresponding target noise signal at four SNR levels (from -5dB to 10dB in 5dB steps) as the test noisy speech. 
Therefore, there are 11,572$\times$4 test noisy utterances for each noise condition.
Three standardized evaluation metrics were used to measure the SE performance: narrow band perceptual evaluation of speech quality ($\text{PESQ}_\text{nb}$)~\cite{PESQ}, short-time objective intelligibility measure (STOI)~\cite{STOI}, and scale-invariant signal-to-distortion ratio (SI-SDR)~\cite{SDR}. 
$\text{PESQ}_\text{nb}$ was designed to evaluate the quality of processed speech, ranging from -0.5 to 4.5.
STOI was designed to compute the speech intelligibility, ranging from 0 to 1. 
SI-SDR was designed to measure the energy ratio between speech and non-speech components. 
For all three metrics, higher scores indicate better performance.

\subsection{Results}
\label{sec:results}
Table~\ref{tab:exp_SE} shows the STOI, $\text{PESQ}_\text{nb}$, and SI-SDR scores of NASTAR and other related SE methods.
\textbf{PTN} represents the original pre-trained SE model. 
\textbf{DAT}$_{\text{one}}$ represents the adaptive SE model trained with DAT using one sample of the test set.
\textbf{DAT}$_{\text{full}}$ represents the adaptive SE model trained with DAT using the full test set. 
To investigate the effect of the noise extractor and the noise retrieval model in our NASTAR model, we prepared the paired data for adaptation in different ways:
(1) only based on the estimated pseudo-noise (termed \textbf{EXTR}, $\alpha$ = 0);
(2) only based on the ground truth of pseudo-noise (termed \textbf{GT}, $\alpha$ = 0);
(3) based on the estimated pseudo-noise and randomly sampled noise signals from DNS-Challenge (termed \textbf{ALL}, $\alpha$ = 0.9);
and (4) only based on the 250 noise signals in relevant-cohort (termed \textbf{RETV}, $\alpha$ = 1).
In addition, we used the second half of the target noise signal to adapt the pre-trained SE model; that is, the adaptation and test noise signal are the same, and $\alpha$ is 0. 
The resulting model is called \textbf{OPT}. 
The oracle performance by \textbf{OPT} represents the upper-bound of NASTAR's performance. 
Furthermore, our test set can be divided into 20 groups based on 5 noise types and 4 SNR levels. 
We performed a dependent t-Test on the paired SE results based on these 20 groups. 
The p-values for NASTAR to outperform \textbf{PTN} in STOI, PESQ$_{\text{nb}}$, and SI-SDR scores are \SI{3.31e-8}, \SI{5.81e-11}, and \SI{3.35e-12}, respectively. 
Based on a standard threshold of 0.05, the improvement of NASTAR over \textbf{PTN} is confirmed to be significant.

\subsubsection{Comparison with other adaptive models}
\label{sec:baseline}
Next, we compare the performance of NASTAR and the adaptive SE model with domain adversarial training (cf. \textbf{DAT}$_{\text{one}}$ and \textbf{DAT}$_{\text{full}}$ in Table~\ref{tab:exp_SE}).  From Table~\ref{tab:exp_SE}, we observe that \textbf{DAT}$_{\text{one}}$ and \textbf{DAT}$_{\text{full}}$ are not always better than \textbf{PTN}, and our NASTAR model consistently outperforms \textbf{DAT}$_{\text{one}}$ and \textbf{DAT}$_{\text{full}}$ in all metrics under all noise conditions.
The main reason for the results is that the help of the auxiliary adversarial training loss cannot effectively improve SE even with complete test data.
In contrast, our NASTAR model is trained to the original SE objective by adjusting the sampling scheme of the existing training data.

\subsubsection{Ablation study}
\label{sec:ablation}
Then, we investigate the effect of the noise extractor and the noise retrieval model in NASTAR. 
From Table~\ref{tab:exp_SE}, we observe that \textbf{EXTR} performs mostly worse than other variants of the NASTAR model.
The results show that due to the estimation error of the noise extractor, the low-quality pseudo-noise is not sufficient to cover the environmental characteristics, and the complement of relevant-cohort is needed.
Comparing NASTAR with \textbf{ALL}, we confirm that adaptive data sampled from relevant-cohort are more useful than randomly sampled data from DNS-Challenge. 
This is because the noise samples in relevant-cohort are closer to the target condition than the randomly sampled data from DNS-Challenge, thus reducing the mismatch between training and test conditions.
\textbf{GT} performs mostly worse than NASTAR and \textbf{RETV}, from which we can see that the sample diversity is not enough to cope with the variation of target noise.
In contrast, the noise samples in relevant-cohort used in NASTAR and \textbf{RETV} are close to the target condition, thus enriching the adaptive data.
Furthermore, \textbf{GT} is not always better than \textbf{EXTR}. 
This result shows that the performance of the noise extractor is acceptable.  
Overall, NASTAR performs the best, followed by \textbf{RETV}. The results confirm the effectiveness of the noise extractor and noise retrieval model in NASTAR.
NASTAR is slightly worse than \textbf{OPT}. There is still room for improvement.

\begin{figure}[h]
    \vspace*{-0.2cm}
    \centering
    \begin{subfigure}[b]{0.22\textwidth}
        \includegraphics[width=\textwidth]{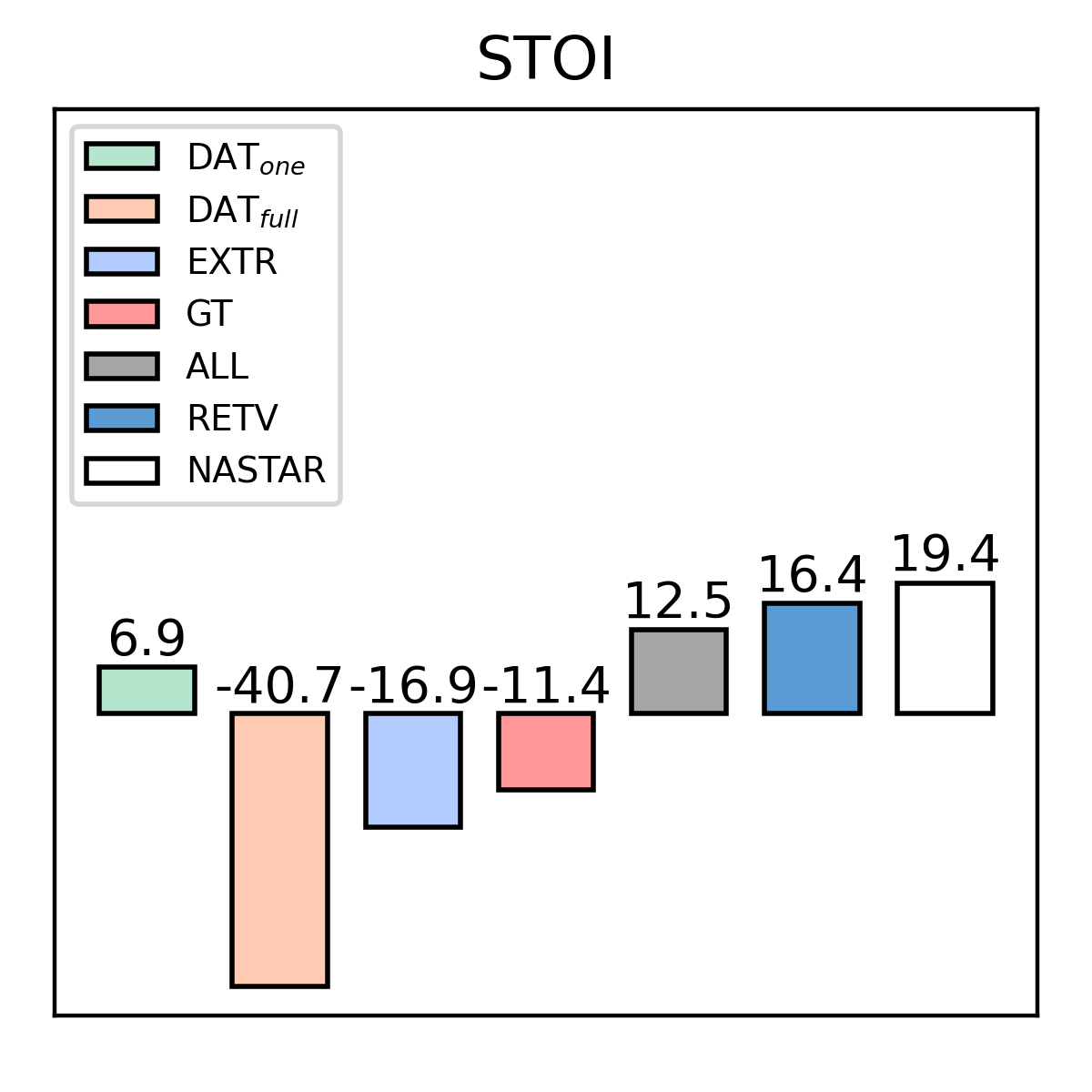}
        \label{fig:stoi}
    \end{subfigure}
    \begin{subfigure}[b]{0.22\textwidth}
        \includegraphics[width=\textwidth]{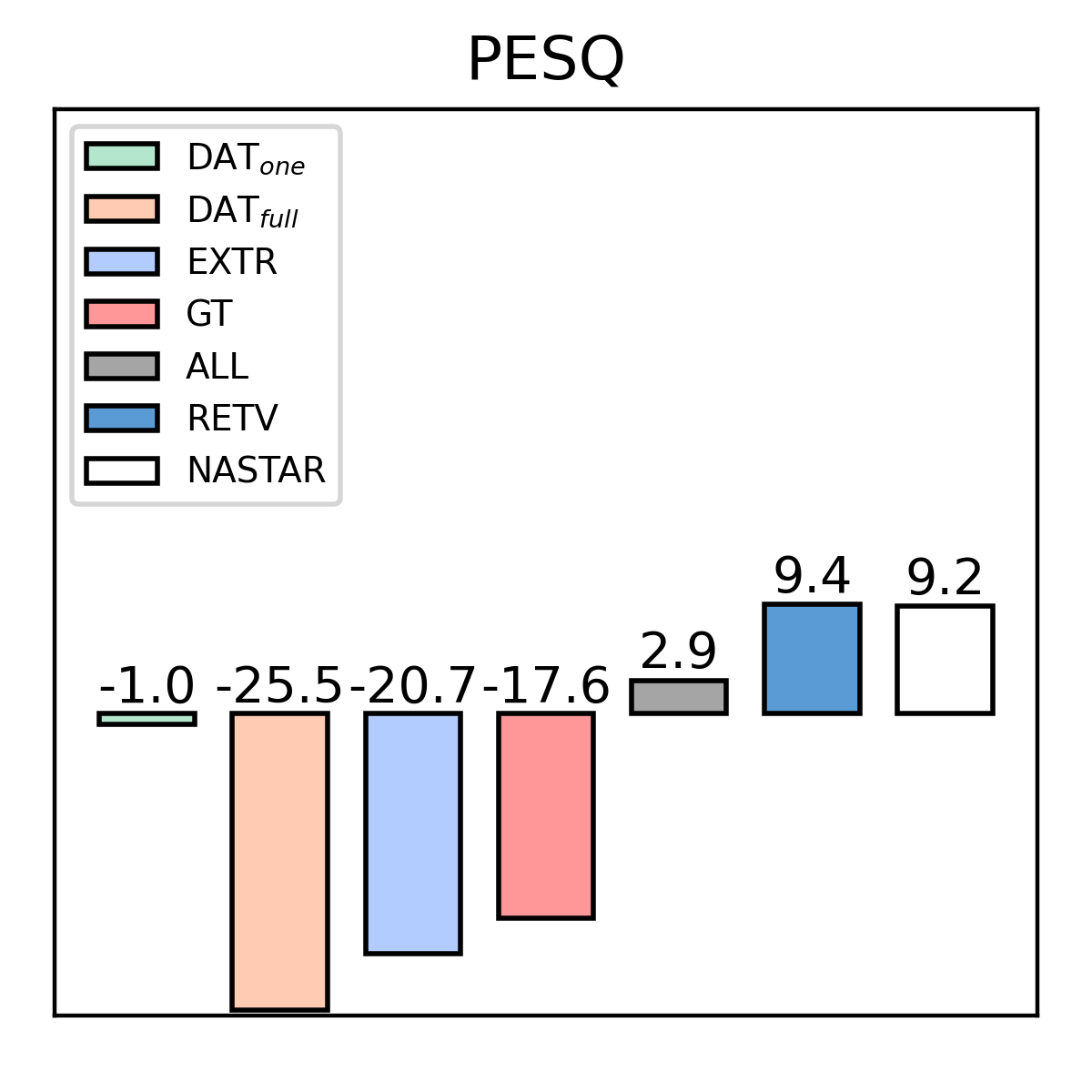}
        \label{fig:pesq}
    \end{subfigure}
    \vspace*{-0.6cm}
    \caption{The average relative improvement rate in STOI and PESQ$_{\text{nb}}$ scores for different SE methods.
    }
    \vspace*{-0.4cm}
    \label{fig:impro}
\end{figure}

\subsubsection{Relative improvement rate}
From Table~\ref{tab:exp_SE}, we note that the STOI improvement of the pre-trained SE model (\textbf{PTN}) over noisy speech is 0.0408 (0.8815 - 0.8407) in \textit{AC/Vacuum}, and the improvement of NASTAR over \textbf{PTN} is 0.0114 (0.8929 - 0.8815) in \textit{AC/Vacuum}. 
We also note that NASTAR has a significant improvement over \textbf{PTN} in terms of $\text{PESQ}_\text{nb}$ and SI-SDR scores. 
As for the other four noise types, we observe the same results that NASTAR consistently outperforms \textbf{PTN} in terms of $\text{PESQ}_\text{nb}$, STOI, and SI-SDR scores.
To further evaluate the improvement brought by adaptation, we define a relative improvement rate as $\frac{s - s_{\text{NOISY}}}
{s_{\text{PTN}} - s_{\text{NOISY}}}$, where $s$ is the metric score obtained by an adaptation method, and $s_{\text{NOISY}}$ is the metric score of the unprocessed noisy speech, and $s_{\text{PTN}}$ is the metric score obtained by the pre-trained SE model (\textbf{PTN}).
The relative improvement rate indicates how much further improvement the adaptation method can provide compared to \textbf{PTN}.
Fig.~\ref{fig:impro} shows the average relative improvement rate in STOI and PESQ$_{\text{nb}}$ scores for NASTAR and other SE systems across five noise types. 
It is clear from the figure that NASTAR reached top performance in both STOI and PESQ$_{\text{nb}}$ compared to other SE methods, confirming the effectiveness of NASTAR. 

\section{Concluding Remarks}
\label{sec:conclusion}

Existing noise adaptation strategies for SE require batches of target noisy speech signals; however, only one-shot can be obtained in most cases.
In this paper, we have proposed a novel NASTAR system to reduce the acoustic mismatch with only one target noisy utterance.
NASTAR uses an accessible sample to simulate adaptation data via a noise extractor and a noise retrieval model.
The noise extractor estimates the target noise, termed pseudo-noise, in the noisy speech sample.
The noise retrieval model retrieves a relevant noise set, termed relevant-cohort, from a noise signal pool according to the given noisy speech sample.
The pseudo-noise and relevant-cohort set are sampled and mixed with the source speech corpus to prepare simulated data for SE model adaptation.
Experimental results show that NASTAR can effectively utilize a single target sample and outperforms several existing methods under different target conditions.
To our best knowledge, NASTAR is the first work to apply resampling to leverage existing source data, which is more precise, efficient, and effective.
Furthermore, our adaptation method can be combined with existing noise-aware and ensemble strategies in various scenarios.
\bibliographystyle{IEEEtran}
\bibliography{main}
\end{document}